\documentclass[twocolumn,twoside]{IEEEtran}

\def\fighome {figures}

\usepackage[mathscr]{eucal}
\usepackage[cmex10]{amsmath}
\usepackage{epsfig,epsf,psfrag,amssymb,amsfonts,latexsym,url,slashbox,graphicx,bm,cite, xcolor,url}
\usepackage[caption=false]{subfig}
\usepackage{fixltx2e}
\usepackage{array}
\usepackage{cases}
\usepackage{verbatim}

\def\tha{\mbox{\tiny th}}

\def\myexp{\mbox{e}}

\def\viz{{viz.,\ \/}}
\def\ie{{i.e.,\ \/}}
\def\eg{{e.g.,\ \/}}
\def\etc{{etc.  }}

\def\as{{a.s.  }}

\def\nn{\nonumber}

\def\qed{\hfill\hbox{${\vcenter{\vbox{
    \hrule height 0.4pt\hbox{\vrule width 0.4pt height 6pt
    \kern5pt\vrule width 0.4pt}\hrule height 0.4pt}}}$}}

\def\tcr{\textcolor{red}}

\definecolor{myred}{rgb}{0.3,0.0,0.7}
\definecolor{dkg}{rgb}{0.1,0.7,0.2}
\definecolor{dkb}{rgb}{0.0,0.2,0.8}

\def\bfzero{{\mathbf{0}}}
\def\bfone{{\mathbf{1}}}

\def\bfy{{\mathbf y}}
 
\def\bfA{{\mathbf A}}

\def\bfI{{\mathbf I}}

\def\bfM{{\mathbf M}}

\def\bfX{{\mathbf X}}
\def\bfY{{\mathbf Y}}
\def\bfZ{{\mathbf Z}}    

\def\Sigmabf{\hbox{$\bf \Sigma$}}

\def\Ac{{\cal A}}
\def\Bc{{\cal B}}
\def\Cc{{\cal C}}

\def\Ec{{\cal E}}

\def\Gc{{\cal G}}
\def\Hc{{\cal H}}

\def\Nc{{\cal N}}

\def\Pc{{\cal P}}

\def\Uc{{\cal U}}
\def\Vc{{\cal V}}

\def\Ebb{{\mathbb E}}

\def\Nbb{{\mathbb N}}

\def\Pbb{{\mathbb P}}

\newcommand{\bprf}{\begin{myproof}}
\newcommand{\eprf}{\end{myproof}}
\newcommand{\bp}{\begin{psfrags}}
\newcommand{\ep}{\end{psfrags}}
\newcommand{\bl}{\begin{lemma}}
\newcommand{\el}{\end{lemma}}
\newcommand{\bt}{\begin{theorem}}
\newcommand{\et}{\end{theorem}}
\newcommand{\bc}{\begin{center}}
\newcommand{\ec}{\end{center}}
\newcommand{\bi}{\begin{itemize}}
\newcommand{\ei}{\end{itemize}}
\newcommand{\ben}{\begin{enumerate}}
\newcommand{\een}{\end{enumerate}}
\newcommand{\bd}{\begin{definition}}
\newcommand{\ed}{\end{definition}}
\def\beq{\vskip .1cm \begin{equation}}
\def\eeq{\end{equation} \vskip .1cm \noindent}
\def\beqn{\vskip .1cm \begin{eqnarray}}
\def\eeqn{\end{eqnarray} \vskip .1cm \noindent}
\def\beqnn{\vskip .01cm \begin{eqnarray*}}
\def\eeqnn{\end{eqnarray*} \vskip .01cm \noindent}
\def\bcase{\vskip .1cm \begin{numcases}}
\def\ecase{\end{numcases} \vskip .1cm \noindent}
\def\bsbcase{\vskip .1cm \begin{subnumcases}}
\def\esbcase{\end{subnumcases} \vskip .1cm \noindent}

\def\defeq{{:=}}

\newtheorem{theorem}{Theorem}
\newtheorem{corollary}{Corollary}
\newtheorem{lemma}{Lemma}
\newtheorem{definition}{Definition}

\newenvironment{myproof}{\noindent{\em Proof:} \hspace*{1em}}{
    \hspace*{\fill} $\Box$ }
\newenvironment{proof_of}[1]{\noindent {\em Proof of #1:}
    \hspace*{1em} }{\hspace*{\fill} $\Box$ }

\def\Pb{p}
\def\LLR{\mbox{LLR}}

\def\V{\mathcal{V}}
\def\E{\mathcal{E}}
\def\nbd{\mathcal{N}}
\def\nnbd{\mbox{nn}}
\def\Deg{\mbox{Deg}}

\def\Rij{R_{ij}}
\def\dist{\mbox{dist}}

\def\nng{\mbox{NNG}}

\def\birootnng{{\mbox{MNNG}}}

\def\snng{{\mbox{\scriptsize NNG}}}

\def\sbirootnng{{\mbox{\scriptsize MNNG}}}

\def\dnng{{\mbox{DNNG}}}
\def\sdnng{{\mbox{\scriptsize DNNG}}}

\title{Detection of Gauss-Markov 
Random Fields with Nearest-Neighbor Dependency}
\author{Animashree Anandkumar, ˜\IEEEmembership{Student~Member,~IEEE,}, 
Lang Tong$^\dagger$\thanks{$^\dagger$Corresponding author.}, 
˜\IEEEmembership{Fellow,~IEEE} and Ananthram Swami, 
˜\IEEEmembership{Fellow,~IEEE}
\thanks{\scriptsize
A.Anandkumar and L.Tong  are with the School of Electrical and 
Computer Engineering, Cornell University, Ithaca, NY 14853.
 Email:{\tt\{aa332@,ltong@ece.\}cornell.edu}}
 \thanks{\scriptsize A. Swami is with the Army Research Laboratory, Adelphi, MD 20783 USA
E-mail: {\tt a.swami@ieee.org}.}
\thanks{\scriptsize This work was supported in part
through the collaborative participation in the Communications and 
Networks Consortium sponsored by the U.~S. Army Research Laboratory 
under the Collaborative Technology Alliance Program, Cooperative 
Agreement DAAD19-01-2-0011 and by the Army Research Office under 
Grant ARO-W911NF-06-1-0346. The U. S. Government is authorized to 
reproduce and distribute reprints for Government purposes 
notwithstanding any copyright notation thereon.} }

\begin{document}
\maketitle
\begin{abstract} The problem of hypothesis testing against independence
for a Gauss-Markov random field (GMRF) is analyzed.  Assuming an 
acyclic dependency graph, an expression for the log-likelihood ratio 
of detection is derived.  Assuming random placement of nodes over a 
large region according to the Poisson or uniform distribution and 
nearest-neighbor dependency graph, the error exponent of the 
Neyman-Pearson detector is derived using large-deviations theory.  
The error exponent is expressed as  a dependency-graph functional 
and the limit is evaluated through a special law of large numbers 
for {\em stabilizing} graph functionals. The exponent is analyzed 
for different values of  the variance ratio and correlation. It is 
found that a more correlated GMRF has a higher exponent at low 
values of the variance ratio whereas the situation is reversed at 
high values of the variance ratio.
\end{abstract}

\vspace{1em}

\noindent

\section{Introduction}\label{section:intro}
\IEEEPARstart{F}or distributed detection, the so-called 
conditionally IID assumption is mathematically convenient and is 
widely assumed in the literature.  The assumption states that 
conditioned on a particular hypothesis, the observations at sensors 
are independent and identically distributed.  In practice, however, 
spatially distributed sensors often observe correlated data, since 
natural spatial signals have stochastic dependence.  Examples of 
correlated signals include measurements from temperature and 
humidity sensors, or from magnetometric sensors tracking a moving 
vehicle. Audio data is also rich in spatial correlations, due to the 
presence of echoes. 

Spatial random signals are typically acausal in contrast to temporal 
signals.  In the literature, the two are usually distinguished by 
referring to acausal signals as random fields (RF) and causal 
signals as random processes (RP).  Random fields are of interest in 
a variety of engineering areas and may represent natural phenomena 
such as the dispersion of atmospheric pollutants, groundwater flow, 
rainfall distribution or the mesoscale circulation of ocean fields 
\cite{Moura&Goswami:97IT}. 

\begin{figure}[t]
\subfloat[a][\scriptsize $\Hc_1$: Gauss-Markov random field with 
nearest-neighbor dependency.]{
\begin{minipage}{1.4in}
\begin{center}
\includegraphics[width = 1.4in]{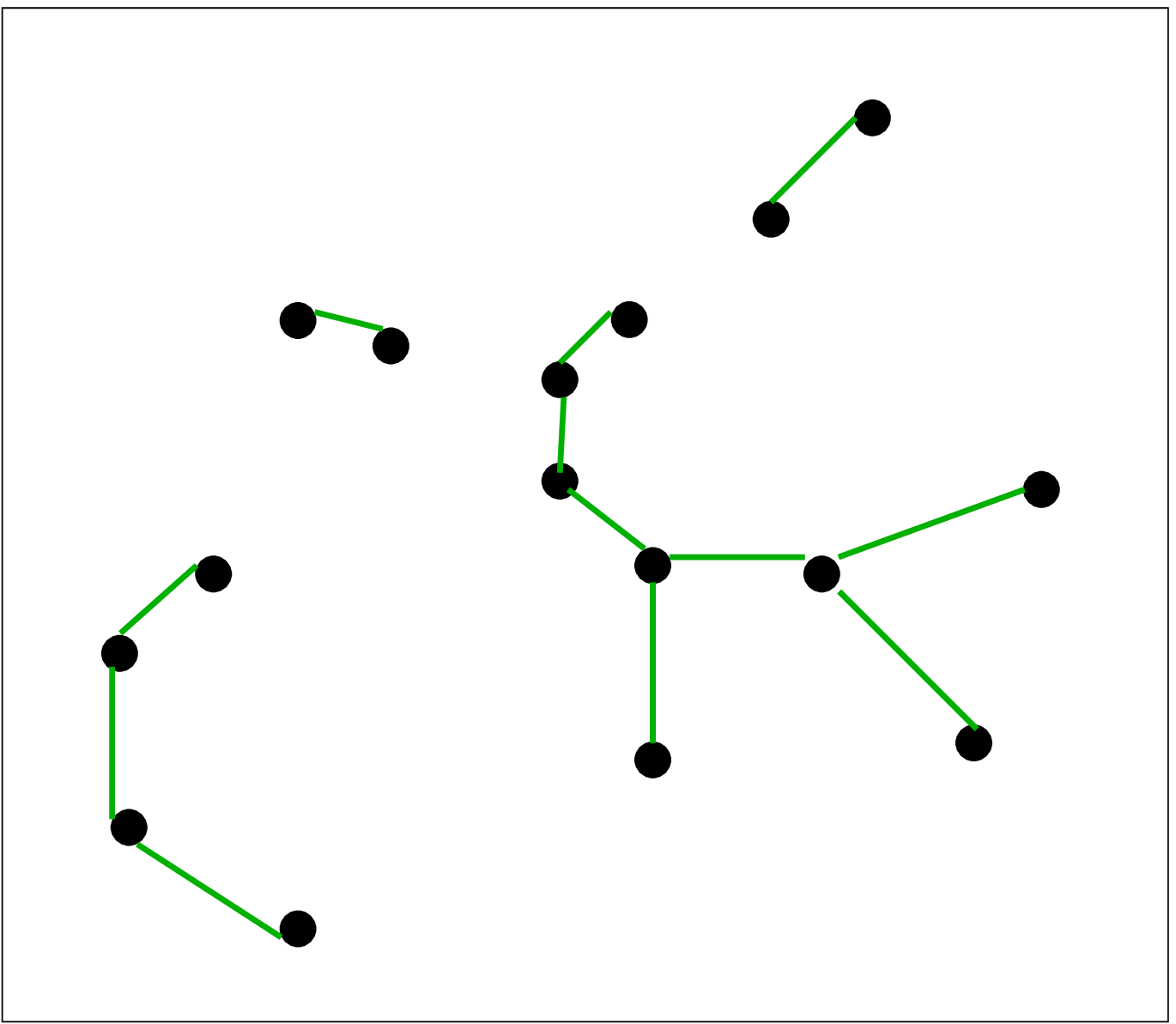} 
\end{center}
\end{minipage}}\hfil 
\subfloat[b][\scriptsize $\Hc_0$ : Independent observations.]{
\begin{minipage}{1.4in}
\begin{center}
\includegraphics[width = 1.4in]{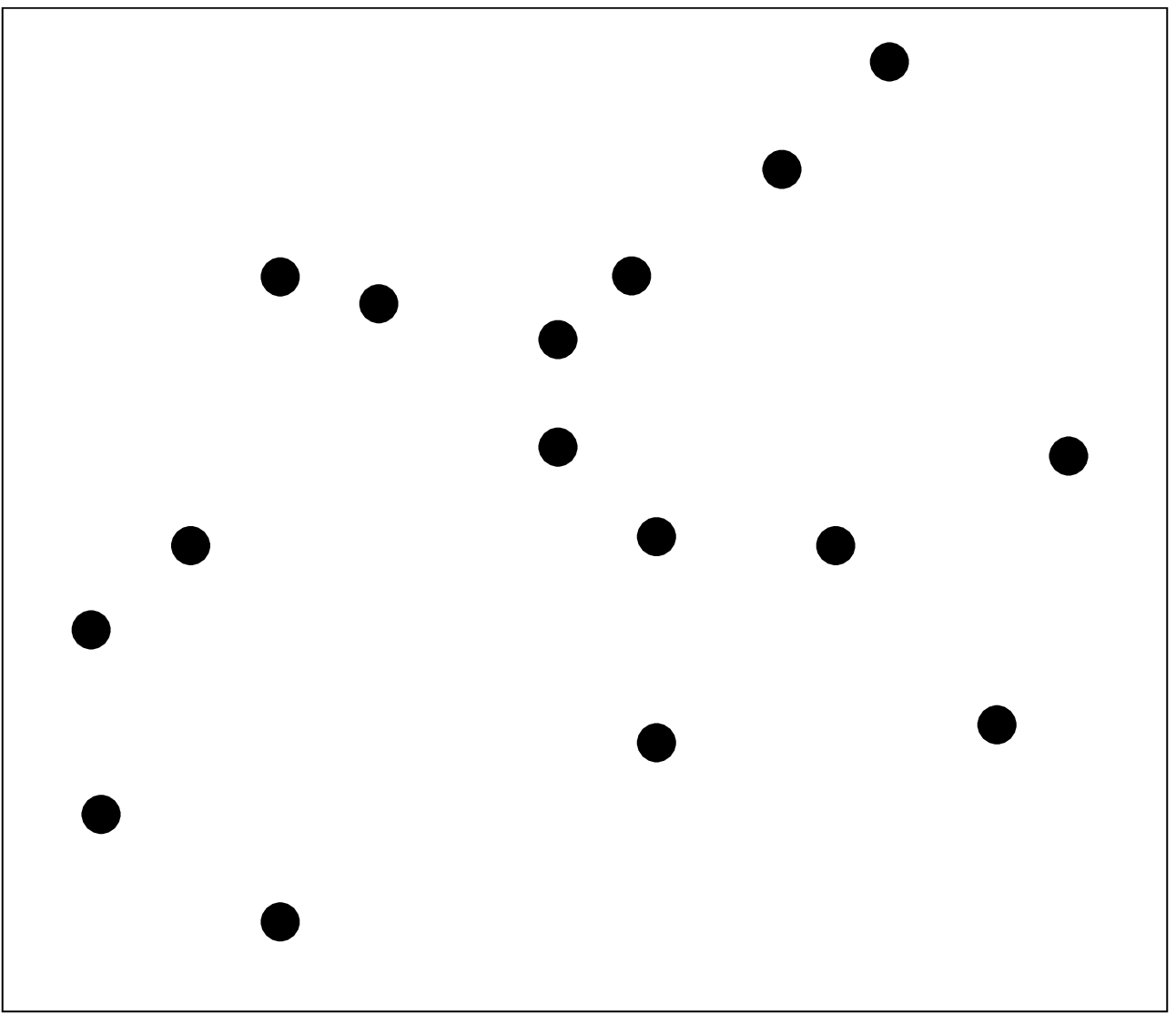}
\end{center}
\end{minipage}}
\caption{Illustration of the hypothesis-testing 
problem.}\label{fig:detection}
\end{figure}

In this paper, we consider the problem of hypothesis testing for 
independence, shown in Fig.\ref{fig:detection}. Specifically, under 
the alternative hypothesis, sensors collect samples from a 
Gauss-Markov random field (GMRF), whereas the samples are 
independent under the null hypothesis. We model the GMRF through a 
graphical approach, in which a dependency graph (DG) specifies the 
stochastic dependence between different sensor observations. This 
dependency graph can have different degrees of sparsity  and can 
even be fully connected.  However typically, spatial interactions 
are based on proximity,  where the  edges are included  according to 
some specified rule  based on the local point configuration 
\cite{Devroye:88math,Penrose&Yukich:01AAP}.  With a regular lattice 
structure (\eg in image processing, Ising model), a fixed set of 
neighbors can be specified in a straight-forward manner 
\cite{Moura&Balram:92IT}.  However, the situation is more 
complicated for arbitrarily placed nodes. In this paper, we consider 
the  nearest-neighbor graph (NNG), which is the  simplest proximity 
graph.    The nearest-neighbor relation has been used in several 
areas of applied science, including the social sciences, geography 
and ecology, where proximity data is often important 
\cite{Cressie:Book,Pace&Zou:00GA}. 




We consider the Neyman-Pearson (NP) formulation, where the detector 
is optimal at a fixed false-alarm probability. But, under this 
formulation, analysis of performance metrics such as error 
probability is intractable for an arbitrary number of observations. 
Hence, we focus on the large-network scenario, where the number of 
observations goes to infinity.  For any positive fixed level of 
false alarm or the type-I error probability, when the mis-detection 
or the type-II error  probability  $P_M(n)$ of the NP detector 
decays exponentially with the sample size $n$, we have the error 
exponent defined by \beq\label{eqn:pm} D 
\defeq -\lim_{n \to \infty} \frac{1}{n}\log P_M(n).\eeq The error 
exponent is an important performance measure since a large exponent 
implies faster decay of error probability with increasing sample 
size. 

Additionally, we assume that the sensors observing the signal field 
are placed i.i.d. according to the uniform or Poisson distribution.  
Since nodes are placed irregularly, it results in a non-stationary  
GMRF (for the definition of stationary GMRF, see \cite[P. 
57]{Rue&Held:book}).  We assume that the number of nodes goes to 
infinity, by way of the coverage area of the nodes going to 
infinity, while keeping the node density fixed. Under this 
formulation, we derive the detection error exponent, 
assuming access to all the observations.   

\subsection{Related Work and Contributions}\label{subsec:related}

The kind of hypothesis testing we consider is called testing for 
independence. In \cite{Ahlswede&Csiszar:86IT,Han&Amari:98IT}, 
problems of this kind are considered with rate constraints on the 
channels and for two sources, using a large number of samples at 
each source.  In this paper, we assume that there are no constraints 
on the channel and that the observations have the correlation 
structure of the GMRF.  Our formulation is different since there is  
a single observation at every sensor, and the number of sensors goes 
to infinity.

GMRF is also known as conditional auto-regression (CAR) in the 
seminal work of Besag \cite{Besag:74JRSS,Besag:75Stat}. They have a 
wide array of applications in fields such as speech recognition, 
natural language processing, coding, geo-statistics, image analysis 
and AI.  The literature is too vast to mention here. For an 
exposition on 
GMRF, see \cite{Rue&Held:book, Guyon:book}. 


Another related problem is the detection of Gauss-Markov random 
processes (GMRP) in Gaussian noise, which is a classical problem 
\cite{Kailath&Poor:98IT}. There is an extensive literature on the 
large-deviations approach to the analysis of  detection of GMRP 
\cite{Donsker&Varadhan:85CMP, Benitz&Bucklew:90IT, Bahr:90IT, 
Bahr&Bucklew:90SP, Barone&Gigli&Piccioni:95IT, Bryc&Smolenski:93SPL, 
Bryc&Dembo:97JTP, Bercu&Gamboa&Rouault:97SPA, Bercu&Rouault:02TPA, 
Chamberland&Veeravalli:04ITWS,Chen:96IT}, but closed-form 
expressions have been derived only for some special cases, \eg  
 \cite{Vajda:90SPA, Luschgy&Rukhin&Vajda:93SPA,Sung&Tong&Poor:06IT}.  
GMRP has been characterized via inversion algorithms for 
block-banded matrices \cite{Kavcic&Moura:00IT,Asif&Moura:05SP}. 
However, these approaches are not amenable to the extension of the 
problem to planar and higher dimensional spaces, since they deal 
with random processes rather than random fields, or to the random 
placement of nodes.

Related to the GMRF, there is an alternative and more restrictive 
approach, known as the spatial auto-regressive model (SAR) and has 
been extensively studied in the field of spatial data-mining. In 
\cite{Pace&Zou:00GA}, this formulation is considered with (directed) 
nearest-neighbor interaction and a closed-form ML estimator of the 
AR spatial parameter is characterized.  We do not consider this 
formulation in this paper.

To our knowledge, large-deviation analysis of the detection of 
acausal non-stationary GMRF has not been treated before. We first 
express the likelihood function of a GMRF with an arbitrary acyclic 
dependency graph, in terms of its covariance matrix. The joint 
distribution can also be derived by expressing it in terms of the 
marginal probability of the nodes and the joint probability at the 
edges of the dependency graph 
\cite{Cowell:book,Wainwright&etal:03IT}.


We consider the detection problem represented in 
Fig.\ref{fig:detection}, under the additional assumptions of 
nearest-neighbor dependency.  We consider the location of the 
sensors as a random point set drawn from uniform or Poisson 
distribution and defined on expanding regions.  This framework 
allows us to exploit recent advances in computational geometry 
\cite{Penrose&Yukich:02AAP,Penrose&Yukich:03AAP}.  By casting the 
error exponent as a limit of the sum of graph functionals, we are 
able to apply the special law of large numbers (LLN) for functionals 
on graphs derived in \cite{Penrose&Yukich:02AAP}.  We obtain the 
final form of the exponent by exploiting some special properties of 
the NNG. We then numerically evaluate the exponent for different 
values of the variance ratio and correlation, for exponential  and 
constant correlation functions.   We conclude that at a fixed node 
density, a more correlated GMRF has a higher exponent at low values 
of variance ratio, whereas the opposite is true at high values of 
variance ratio.  


\subsection{Notation and Organization}%

Vectors and matrices are written in boldface.  Random variables are 
in capital letters, random processes and random fields in boldface 
capitals and sets in calligraphic font. For the matrix $\bfA 
=[A(i,j)]$,  $A(i,j)$ denotes the  element in the $i^{\tha}$ row and 
$j^{\tha}$ column and $|\bfA|$ its determinant.  For  sets $\Ac$ and 
$\Bc$, let $\Ac  \backslash \Bc =\{ i: i\in \Ac, i \notin \Bc\} $ 
and let $|\cdot|$ denote cardinality.

An undirected graph $\Gc$ is a tuple  $\Gc=(\V, \E)$ where  
$\V=\{1,2,\ldots,n\}$ is the vertex\footnote{We consider the terms 
node, vertex and sensor interchangeable.} set  and $\E =\{(i,j),\, 
i,j \in \V, i\neq j\}$ is the edge set.  When $i$ and $j$ have an 
edge between them, $i$ and $j$ are neighbors denoted by $i \sim j$ 
(otherwise it is $i \nsim j$). For a directed graph, we denote the 
edges by $ \E =\{<i,j>,\, i,j \in \V, i\neq j\}$, where the 
direction of the edge is from $i$ to $j$. The neighborhood function 
of a node $i$ is the set  of all other nodes having an edge with it, 
\ie \beq \nbd(i) = \{j \in \V : j \neq i, (i,j) \in \E\}.\eeq  The 
number of neighbors of a node $i$ is called its degree, denoted by 
$\Deg(i)$.  A node with a single edge \ie its degree is $1$ is known 
as a leaf and the corresponding edge as a leaf edge, otherwise it is 
known as an internal or interior edge. Let $\dist(i,j)$ be the 
Euclidean distance between any two nodes. Let $R_{ij}$ denote the 
(random) Euclidean edge-length  of $(i,j)$ in graph $\Gc=(\V,\E)$, 
\beq R_{ij}=\dist(i,j),\quad \forall\,(i,j)\in 
\E.\label{eqn:Rij}\eeq

Our paper is organized as follows.  We provide a description of the 
GMRF in section~\ref{sec:gmrf}, focusing on the acyclic dependency 
graph in section~\ref{subsec:acyclic} and providing an expression 
for the likelihood function in section~\ref{subsec:potential}. We 
define the hypothesis-testing problem in section~\ref{sec:problem} 
and specify additional assumption on the covariance matrix of the 
GMRF in section~\ref{subsec:corr}. In section~\ref{subsec:nng}, we 
assume additionally that the dependency graph is the 
nearest-neighbor graph. We provide an expression for the 
log-likelihood ratio in section~\ref{subsec:llr}. We define the 
error exponent under the Neyman-Pearson formulation in 
section~\ref{sec:error_exponent} and specify  the random placement 
of nodes in section~\ref{subsec:random_sets}. In 
section~\ref{sec:results} we evaluate the error exponent, expressing 
it as a graph functional in section~\ref{subsec:graph}, applying the 
LLN for graphs in section~\ref{subsec:lln}, and providing an 
explicit form for NNG in section~\ref{subsec:explicit_results}. We 
provide numerical results for the exponent in 
section~\ref{subsec:numerical}, and section~\ref{sec:conclusion} 
concludes the paper. 

\section{Gauss-Markov Random Field}\label{sec:gmrf}


%

A GMRF, in addition to being a Gaussian random field, satisfies 
special conditional independence properties. A simple example is the 
first-order auto-regressive process, where the conditional 
independence of the observations is based on causality.  However, a 
spatial random field has a far richer set of conditional 
independencies, requiring a more general definition \cite[P. 
21]{Rue&Held:book}.  
\begin{definition}[GMRF]
Given a point set  $\V=\{1,\ldots,n\}$, $\bfY_\V = \{Y_i : i \in 
\V\}$ is a GMRF with an (undirected) dependency graph $\Gc(\V,\E)$  
if $ \bfY_\V$ is a Gaussian random field, and $\forall i,j \in \V$, 
$Y_i$ and $Y_j$ are conditionally independent given observations at 
all other nodes if $i$ and $j$ are not neighbors, \ie  \beq Y_i 
\perp Y_j| \bfY_{-ij} \iff i \nsim j,\,\,\forall i,j \in \V, i\neq 
j,\eeq where $\perp$ denotes conditional independence and 
$\bfY_{-ij} 
\defeq(Y_k: k \in \V, k \neq i,j)$.\end{definition} 

A common approach to formulating a GMRF is to specify the dependency 
graph through a neighborhood rule and then to specify the 
correlation function between these neighbors. Thus, in a GMRF, local 
characteristics completely determine the joint distribution of the 
Gaussian field.  


\begin{figure}[t]
\hspace{-0.35in} \subfloat[a][A labeled simple undirected graph.]{
\begin{minipage}{1.4in}
\centering
\begin{psfrags}
\includegraphics[width =1.4in,height = 1.1in]{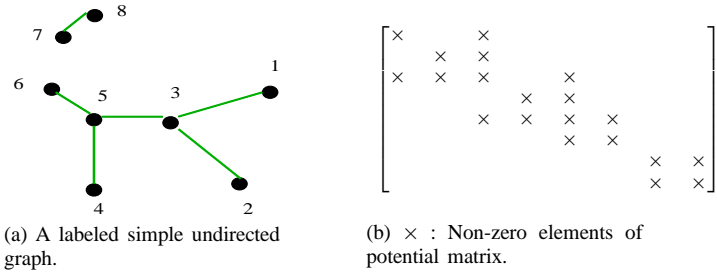}\end{psfrags}
\end{minipage}}
\hfil\subfloat[b][$\times$ : Non-zero elements of potential 
matrix.]{
\begin{minipage}{1.4in}
{\scriptsize \[\begin{bmatrix} \times & & \times & & & & & \\ & 
\times & \times & & & & & \\\times & \times & \times & & \times& & & 
\\ & & &\times & \times & & & \\ & & \times & \times & \times & 
\times & & 
\\ & & & & \times & \times & & \\ & & & & & & \times & \times \\ & & 
& & & & \times & \times 
\end{bmatrix}\]} 
\end{minipage}}
\caption{Dependency graph and potential matrix of a 
GMRF.}\label{fig:labelled_graph}
\end{figure}

The following Markov properties are equivalent in a GMRF:\ben
\item pairwise-Markov property \beq Y_i \perp Y_j| 
\bfY_{-ij} \iff (i, j) \notin \Ec.\eeq

\item local-Markov property \beq Y_i \perp \bfY_{-(i, \nbd(i))} | 
\bfY_{\nbd(i)},\label{eqn:local_markov}\eeq

\item global-Markov property \beq\bfY_{\Ac} \perp \bfY_{\Bc} | \bfY_{\Cc},\label{eqn:global_markov}\eeq 
for disjoint sets $\Ac$, $\Bc$, and $\Cc$, with $\Ac$ and $\Bc$ 
non-empty, where the set $\Cc$ separates $\Ac$ and $\Bc$ \ie  on 
removing the nodes in $\Cc$ from the graph, nodes in $\Ac$ are no 
longer connected to the nodes in $\Bc$.\een

Thus, in (\ref{eqn:local_markov}), the local-Markov property states 
that the conditional distribution at a node in the DG given the 
observations at its neighbors is independent of the rest of the 
network. By the global-Markov property in (\ref{eqn:global_markov}), 
all the connected components of a dependency graph are independent.  
As an illustration, in Fig.\ref{fig:labelled_graph} we have 
$Y_6\perp Y_7$ given the rest of network, $Y_1\perp Y_2 | Y_3$, and 
so on. 

\section{Acyclic Dependency Graph}\label{subsec:acyclic}

A special case of the dependency graph is an acyclic or a cycle-free 
graph. Here, the neighbors of a node are not themselves neighbors. 
The joint distribution is somewhat easier to evaluate in this case. 
We note that an acyclic graph with at least one edge, always has a 
leaf \ie  it has a node with degree $1$ and has utmost $n-1$ edges 
in a $n$-node graph.
  
The covariance matrix $\Sigmabf$ of a GMRF satisfies some special 
properties. For instance, consider the cross covariance between the 
neighbors of a node, \ie nodes that are two hops away in an acyclic 
DG.  By the global-Markov property we have\footnote{For $X,Y$ 
jointly zero mean Gaussian,  $\Ebb(\bfX|\bfy) =\Sigmabf_{xy} 
\Sigmabf_{yy}^{-1}\bfy$.}, assuming $\Sigma(i,i)>0$, for $ i \in \V, 
\, \Deg(i) \geq 2,\,j,k \in \nbd(i), j \neq k$,  \beq \Sigma(j,k) = 
\frac{\Sigma(i,j)\Sigma(i,k)}{\Sigma(i,i)}.\label{eqn:nonneighbor_gen}\eeq 
For example, in Fig.\ref{fig:labelled_graph}, \beq \Sigma(1,2) = 
\frac{\Sigma(1,3)\Sigma(2,3)}{\Sigma(3,3)}.\eeq We can similarly 
find an expression for the covariance between any two nodes of the 
GMRF. Thus, the covariance matrix of a GMRF with acyclic dependency 
can be expressed solely in terms of the auto covariance of the nodes 
and the cross covariance between the neighbors of the dependency 
graph.


\subsection{Potential Matrix}\label{subsec:potential}

The inverse of the covariance matrix  of a non-degenerate GMRF (\ie 
with a positive-definite covariance matrix) is known as the 
potential matrix or the precision matrix or the information matrix. 
The non-zero elements of the potential matrix $\bfA$ are in one to 
one correspondence with the edges of its graph $\Gc(\V,\E)$ 
\cite[Theorem 2.2]{Rue&Held:book} in the sense that \beq i \nsim j 
\iff A(i,j) =0, \,\forall i,j \in \V, i\neq j,\eeq and is 
illustrated in Fig.\ref{fig:labelled_graph}. 

This simple correspondence between the conditional independence of 
the GMRF and the zero structure of its potential matrix is not 
evident in the covariance matrix, which is generally a completely 
dense matrix.  Therefore,    it is easier to evaluate the joint 
distribution of the GMRF through the potential matrix.  In practice, 
however, estimates of the covariance matrix are easier to obtain 
through the empirical observations.  Therefore,    it is desirable 
to have the joint distribution in terms of coefficients of the 
covariance matrix. Thus, an explicit expression between the 
coefficients of the covariance and the potential matrix is needed. 
We provide such an expression and also obtain the determinant of the 
potential matrix in the theorem below. 

\begin{theorem}[Elements \& Determinant of Potential Matrix]\label{thm:potential&determinant}
The elements of the potential matrix  $\bfA 
\defeq\Sigmabf^{-1}$,  for a positive-definite
covariance matrix $\Sigmabf$ and acyclic dependency graph 
$\Gc(\V,\E)$, are \beqn A(i,i) &=& \frac{1}{\Sigma(i,i)} \Bigl(1 
+\sum_{j \in \nbd(i)} \frac{\Sigma(i,j)^2}{\Sigma(i,i) \Sigma(j,j) - 
\Sigma(i,j)^2}\Bigr),\nn\\A(i,j) &=& \left\{\begin{array}{c c} 
\dfrac{-\Sigma(i,j)}{\Sigma(i,i) \Sigma(j,j) - \Sigma(i,j)^2}& 
\mbox{if}\,\,  i \sim j,\\ 0 & \mbox{o.w.}\end{array} 
\right.\label{eqn:aij_general}\eeqn The determinant of the potential 
matrix of  $\bfA$ is given by  \beq |\bfA|=\frac{1}{|\Sigmabf|} = 
\frac{\prod_{i \in \V} \Sigma(i,i)^{{\scriptsize \Deg(i)} 
-1}}{\prod\limits_{\substack{i \sim j\\i<j}} [\Sigma(i,i)\Sigma(j,j) 
- \Sigma(i,j)^2]}.\label{eqn:determinant_general}\eeq 
\end{theorem}

\bprf The proof is based on acyclicity of dependency graph. See 
Appendix~\ref{proof:potential&determinant}.\eprf

\section{Hypothesis-Testing Problem}\label{sec:problem}


\noindent Let $\V = \{1,\ldots,n\}$ be a set of  $n$ nodes on the 
plane and let $\bfY_n$ be the random vector of observation samples 
$Y_i, i \in \V$, \beq \bfY_n 
\defeq [Y_1,\ldots,Y_n]^T.\eeq The hypothesis-testing problem is as follows 
(also see Fig.\ref{fig:detection}), \beq\Hc_0 : \bfY_n \sim 
\Nc(\bfzero, \sigma_0^2 \bfI) \quad \mbox{vs.} \quad \Hc_1: \bfY_n 
\sim \Nc(\bfzero,\Sigmabf_{1,\V}),\label{eqn:detection}\eeq where 
$\Sigmabf_{1,\V}$ is a positive-definite covariance matrix  under 
the alternative hypothesis  and is dependent on the configuration of 
nodes in $\V$ and $\sigma_0^2>0$ is the uniform variance under the 
null hypothesis.


The optimal decision-rule under both NP and Bayesian formulations is 
a threshold test based on the log-likelihood ratio (LLR). Let 
$\Pb[\bfY_n|\V;\Hc_j]$ be the conditional PDF of the observations 
given the point set $\V$ under hypothesis $j$. The LLR given by 
\beqn \LLR(\bfY_n, \V)\!\!\!\! &\defeq& \!\!\!\! 
\log\frac{\Pb[\bfY_n,\V;\Hc_0]}{\Pb[\bfY_n,\V;\Hc_1]}= 
\log\frac{\Pb[\bfY_n;\Hc_0]}{\Pb[\bfY_n|\V;\Hc_1]},\label{eqn:llr_def}\\\!\!\!\! 
&=& \!\!\!\! \!\!\frac{1}{2}\Bigl(\log\frac{|\Sigmabf_{1,\V}| 
}{|\sigma_0^2 \bfI| } + \bfY_n^T[\Sigmabf^{-1}_{1,\V}- (\sigma_0^2 
\bfI)^{-1}]\bfY_n\Bigr),\nn  \eeqn where in (\ref{eqn:llr_def}), we 
have used the fact that the sensor observations are independent of 
$\V$ under $\Hc_0$.  

\subsection{Covariance Matrix of GMRF}\label{subsec:corr}

We make additional assumption on the structure of the covariance 
matrix $\Sigmabf_{1,\V}$ of the GMRF under $\Hc_1$ \viz that the 
nodes have the same measurement variance  for any node configuration 
$\V$, \ie  \beq \Sigma_{1,\V}(i,i) 
\defeq \sigma_1^2>0,\,\, i=1,\ldots,n.\label{eqn:selfcorr}\eeq  
We denote the ratio between the variances under the alternative and 
the null hypothesis at each node by \beq K
\defeq \frac{\sigma_1^2}{\sigma_0^2}.\label{eqn:K}\eeq

We also assume that under $\Hc_1$, the amount of correlation between 
the neighbors $i,j$ of the dependency graph is specified by an 
arbitrary function $g$, which has the Euclidean edge length $R_{ij}$ 
as its argument.  From (\ref{eqn:selfcorr}), we have \beq 
g(R_{ij})\defeq\frac{\Sigma_{1,\V}(i,j)}{\sigma_1^2}<1,\,\, 
\forall\,\, (i,j) \in \E.\label{eqn:corr}\eeq The correlation 
function $g$ is required to satisfy some regularity conditions, 
which will be stated in Lemma~\ref{lemma:conditions}. In 
general, $g$ is a monotonically non-increasing function of 
the edge length, since amount of correlation usually decays as nodes 
become farther apart. Moreover, $g(0) = M <1$, or the so-called 
nugget effect, according to geo-statistics literature 
\cite{Moller:book,Jaksa&etal:06iae}.  It has been observed  in 
mining applications, where the micro-scale variation is assumed to 
be caused by the existence of small nuggets of the enriched ore.  
Many other ecological phenomena such as soil bacteria population 
\cite{Grundmann:00}, aquatic population \cite{Barange97} \etc also 
exhibit this behavior. Note that the presence of nugget effect has 
the same effect on correlation as imposing an exclusion region on 
how near two nodes can be placed. However, for such an exclusion 
constraint to hold, we need more complicated node placement 
distributions than the uniform or Poisson assumption. Although such 
distributions can be handled in principle, they are not analytically 
tractable.

Some examples of the correlation function are 
\[g(R) =M\myexp^{-aR}, \, g(R)= \frac{M}{1+R^a},\,\,a\geq0,0\leq M < 
1.\]Note that these conditions together with an acyclic dependency 
graph $\Gc$ assure that the covariance matrix $\Sigmabf_{1,\V}$ is 
positive definite. This is because $\forall$ $i,j \in \V$, 
\[\Sigma_{1,\V}(i,i)\Sigma_{1,\V}(j,j) - \Sigma_{1,\V}^2(i,j) = 
\sigma_1^4[1-\!\!\!\!\!\!\!\!\prod_{(k,l)\in  \mbox{\scriptsize 
path}(i,j)} \!\!\!\!g^2(R_{kl})]>0,\] where $\mbox{path}(i,j)$ is 
the unique edge-path between $i$ and $j$ in graph $\Gc$ if it 
exists.  From Theorem~\ref{thm:potential&determinant}, we have 
$|\Sigmabf|>0$.






\subsection{Nearest-Neighbor Graph}\label{subsec:nng}

We assume the dependency graph to be the nearest-neighbor graph. The 
nearest-neighbor function of a node $i \in \V$, is defined as, \beq 
\nnbd(i) \defeq \arg\min_{j \in \V, j\neq i } 
\dist(i,j),\label{eqn:nnbd}\eeq where $\dist(\cdot,\cdot)$ is the 
Euclidean distance. The inter-point distances are unique with 
probability 1, for uniform and Poisson point sets under 
consideration here. Therefore, $\nnbd(i)$ is a well-defined function 
almost surely. The nearest-neighbor (undirected) graph $\Gc(\V,\E)$ 
is given by \beq (i,j) \in \E \iff i = \nnbd(j) \,\,\mbox{or}\,\, j 
= \nnbd(i).\eeq  NNG has a number of important properties. It is 
acyclic with a maximum\footnote{The  node degree is finite for NNG 
in any dimension and is called the kissing number 
\cite{zong:98math}.} node degree of $6$  
\cite{Eppstein&Paterson&Yao:97}. 

\begin{figure}[t]
\begin{center}
\begin{psfrags}
\psfrag{Biroots of}[l]{\scriptsize Biroots of}\psfrag{directed 
NNG}[l]{\scriptsize of directed NNG}\psfrag{Directed 
NNG}[l]{\scriptsize Directed NNG}\psfrag{Undirected 
NNG}[l]{\scriptsize Undirected NNG}\includegraphics[height = 
1.1in]{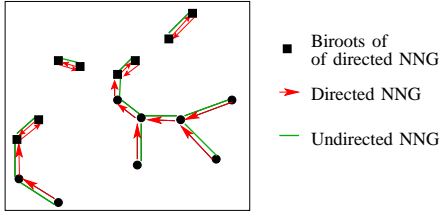}\end{psfrags} 
\caption{\small Directed \& undirected versions of nearest-neighbor 
graph. [ No. of undirected edges = No. of directed edges - $0.5 
\times$ No. of biroots.] }\label{fig:directed_graph} 
\end{center}
\end{figure}

In section~\ref{subsec:explicit_results}, it turns out that we need 
to analyze the directed NNG, in order to obtain the final form of 
the error exponent. We now mention some of its special properties. 
The directed NNG $\Gc'(\V,\Ec')$ is defined by \beq \E' = \{ 
<i,\nnbd(i)>, i\in \V\},\eeq For a directed NNG with at least two 
nodes, each connected component contains exactly one 2-cycle. This 
is known as the biroot of the component 
\cite{Eppstein&Paterson&Yao:97}. See Fig.\ref{fig:directed_graph}. 
Also note, the directed NNG counts the edges from these biroots 
twice, while the undirected version counts only once. 

\subsection{Expression for Log-Likelihood Ratio}\label{subsec:llr}

Since the NNG is acyclic, equations 
(\ref{eqn:aij_general}-\ref{eqn:determinant_general}) are valid. We 
incorporate additional assumptions 
(\ref{eqn:selfcorr}-\ref{eqn:corr}) in the theorem to obtain the LLR 
for detection.

%

\begin{theorem}[Log-Likelihood Ratio]\label{thm:LLR} Under the assumptions 
(\ref{eqn:selfcorr}-\ref{eqn:corr}), the log-likelihood ratio in 
(\ref{eqn:llr_def}) for the hypothesis-testing problem in 
(\ref{eqn:detection}), given an arbitrary point set $\V=\{1, \ldots, 
n\}$, is
\end{theorem} \beqn \LLR(\bfY_n, \V)&{}={}&n\log\frac{\sigma_1}{\sigma_0}+ \frac{1}{2}\biggl[\sum_{i \in 
\Vc}\Bigl(\frac{1}{\sigma_1^2}-\frac{1}{\sigma_0^2}\Bigr)Y_i^2 \nn\\ 
&&{+}\:\sum_{\substack{(i,j) \in \E\\i<j}}\Bigl\{ \log[1-g^2(\Rij)] 
\nn 
\\&&{+}\:\frac{g^2(\Rij)}{1-g^2(\Rij)}\frac{Y_i^2+ Y_j^2}{\sigma_1^2}\nn 
\\&&{-}\:\frac{2 g(\Rij)}{1-g^2(\Rij)}\frac{ Y_i 
Y_j}{\sigma_1^2}\Bigr\}\biggr],\label{eqn:LLR}\eeqn {\em where 
$\Rij$ is the Euclidean edge length of $(i,j) \in \E$, that depends 
on the configuration of $\V$. The condition $i<j$ ensures that every 
edge is counted only once.}


Theorem~\ref{thm:LLR} gives a closed-form expression for the 
log-likelihood ratio, in terms of the edges of the nearest-neighbor 
dependency graph of the GMRF.  Note  in (\ref{eqn:LLR}), the 
cross-terms are only between the neighbors of the dependency graph, 
which can be exploited to yield explicit data-fusion and routing 
schemes \cite{Anandkumar&etal:08INFOCOM}.

\section{Neyman-Pearson Error Exponent}\label{sec:error_exponent}


The {\em spectrum} of the log-likelihood ratio is defined as the 
distribution of the normalized LLR evaluated under the null 
hypothesis. In \cite[Theorem 1]{Chen:96IT}, it is proven that for 
Neyman-Pearson detection under a fixed type-I error bound 
\footnote{The generalization to an exponential type-I error  bound 
 \cite{Chen:96IT, Han:00IT} is not tractable since  a closed-form 
cumulative distribution of the LLR is needed.}, the LLR spectrum   
can fully characterize the  type-II error exponent of the 
hypothesis-testing system  and is independent of the type-I  bound.

A special case of this result is when the LLR spectrum  converges 
almost surely (a.s) to a constant $D$ 
\[\frac{1}{n}\LLR(\bfY_n, \V)= \frac{1}{n} \log\frac{\Pb[\bfY_n;\Hc_0]}{\Pb[\bfY_n|\V;\Hc_1]} 
\overset{\mbox{a.s.}}{\to} D, \quad \mbox{under}\,\,  \Hc_0.\] In 
this case,   the NP type-II error exponent   is given by the above 
constant $D$. In other words,     the error exponent $D$ of NP 
detection in (\ref{eqn:pm}) is  \beq D 
\defeq \lim_{n \to \infty}\frac{1}{n} 
\log\frac{\Pb[\bfY_n;\Hc_0]}{\Pb[\bfY_n|\V;\Hc_1]},\quad  
\mbox{under}\,\, \Hc_0,\label{eqn:logp} \eeq where $\lim$ denotes 
the almost-sure limit, assuming it exists. Note that when $\bfY_n$ 
are i.i.d. conditioned  under either $\Hc_0$ or $\Hc_1$, the result 
reduces to the Stein's lemma \cite[Theorem 
12.8.1]{Cover&Thomas:Book} and the limit in (\ref{eqn:logp}) to the 
Kullback-Leibler distance. 

\subsection{Random Point Sets}\label{subsec:random_sets}

\begin{figure}[t]
\bc\bp\psfrag{Fusion center}[l]{\scriptsize Fusion center}
\includegraphics[height = 
1.1in]{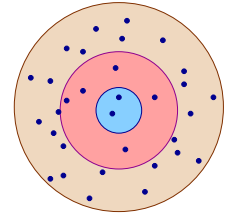}\ep \ec\caption{Illustration 
of point process $\Uc_{n,\lambda}$ or $\Pc_{n,\lambda}$: $n$ nodes 
distributed i.i.d. uniform or Poisson in regular Borel regions (such 
as squares or circles) of area $\frac{n}{\lambda}$ with constant 
density $\lambda$. For error exponent, we consider $n\to 
\infty$.}\label{fig:density}\end{figure}

It is intractable to evaluate the error exponent $D$ in 
(\ref{eqn:logp}) for an arbitrary point set.  Therefore,   we assume 
that the nodes are placed randomly, according to a point process 
defined on expanding regions. We consider two related point 
processes : the Poisson process and the binomial point process on a 
large region, which we define below.

\bd[Poisson and Binomial Processes \cite{Moller2:book}] Let 
$(\Bc_n)_{n\geq1}$ denote a sequence of squares or 
circles\footnote{The results hold for regular Borel sets under some 
conditions \cite[P. 1007]{Penrose&Yukich:01AAP}.}  of area 
$\frac{n}{\lambda}$, centered at the origin, for any  $\lambda>0$.  
A binomial point process on $\Bc_n$, denoted by  $\Uc_{n,\lambda}$,  
consists of $n$ points distributed i.i.d. uniformly on  $\Bc_n$. A 
homogeneous Poisson process of intensity $\lambda$ 
  on $\Bc_n$, denoted by $\Pc_{n,\lambda}$, satisfies the following 
properties:\ben \item for any set $\Ac \subset \Bc_n$ with area $A$, 
the number of points in $\Ac$ is Poisson distributed with mean 
$\lambda A$, \item for any $n \in \Nbb$ and $\Ac \subset \Bc_n$ with 
area $A>0$, conditioned on $n$ number of points in $\Ac$, the point 
process on $\Ac$ is a binomial process. \een  \ed

We are interested in evaluating the error exponent under both the
binomial or Poisson point processes, when the mean number of nodes 
goes to infinity, with fixed node density, \ie $n \to \infty$ with 
$\lambda$ fixed.  



\section{Closed-Form Error Exponent}\label{sec:results}

\subsection{Error Exponent as a Graph 
Functional}\label{subsec:graph}

In order to derive the error exponent, we cast the error exponent as 
the limit of sum of node and edge functionals of the dependency 
graph of a {\em marked} point set in the lemma below.  This 
formulation is required in order to apply the law of large numbers 
for graph functionals.

\bl[$D$ as a Graph Functional]\label{lemma:D}  Given the marked 
point set $\V$ drawn from the binomial process $\Uc_{n,\lambda}$ or 
the Poisson process $\Pc_{n,\lambda}$, with marking variable $Y_i 
\overset{i.i.d.}\sim \Nc(0,\sigma_0^2)$, the error exponent  $D$ in 
(\ref{eqn:logp}) is given by the limit of sum of edge and node 
functionals of the nearest-neighbor graph as \el \beqn D \!\!\!\! 
&{}={}&\!\!\!\!\log\frac{\sigma_1}{\sigma_0}+ \lim_{n \to \infty} 
\frac{1}{2 n}\biggl[\sum_{i \in 
\Vc}\Bigl(\frac{1}{\sigma_1^2}-\frac{1}{\sigma_0^2}\Bigr)Y_i^2 \nn\\ 
\!\!\!\!&&{+}\:\!\!\!\!\sum_{\substack{(i,j) \in \E\\i<j}}\Bigl\{ 
\log[1-g^2(\Rij)] + \frac{g^2(\Rij)}{1-g^2(\Rij)}\frac{Y_i^2+ 
Y_j^2}{\sigma_1^2}\nn 
\\ \!\!&&{-}\: \frac{2 g(\Rij)}{1-g^2(\Rij)}\frac{ Y_i 
Y_j}{\sigma_1^2}\Bigr\}\biggr], \, Y_i \overset{i.i.d.}\sim 
\Nc(0,\sigma_0^2),\label{eqn:D}\eeqn {\em where $\Rij$ is the 
(random) Euclidean edge length of $(i,j) \in \E$, that depends on 
the underlying point process. The condition $i<j$ ensures that every 
edge is counted only once.}

\noindent Proof : Substitute (\ref{eqn:LLR}) in 
(\ref{eqn:logp}).\qed

In the lemma above, the point set forming the graph is drawn from a 
marked binomial or Poisson point process, with the marking variable 
$Y_i\overset{i.i.d.}\sim \Nc(0,\sigma_0^2)$.  This is because 
evaluating the error exponent (\ref{eqn:logp}) under $\Hc_0$ implies 
that the sensor observations $Y_i$ are i.i.d. and independent of the 
locations of the nodes, and therefore can be viewed as a marking 
process. 

\subsection{Law of Large Numbers for Graph Functionals}\label{subsec:lln}
\begin{figure}[t]\bc\bp\psfrag{n to
infinity}[l]{\scriptsize $n \to 
\infty$}\psfrag{Origin}[l]{\scriptsize Origin} \psfrag{Normalized 
sum of edge weights}[l]{\scriptsize Normalized sum of edge 
weights}\psfrag{Expectation of edges}[l]{\scriptsize Expectation of 
edges}\psfrag{of origin of Poisson process}[l]{\scriptsize of origin 
of Poisson process}\psfrag{edge sum}[l]{\scriptsize \tcr{ 
$\frac{\sum\limits_{e \in \Gc(\V)} \Phi(R_e)}{n}$}}\psfrag{edge sum 
poisson}[l]{\scriptsize \tcr{$\frac{1}{2}\Ebb 
\sum\limits_{\substack{\bfX:\bfX \in \Pc_{\lambda}\\ 
(\bfX,\bfzero)\in \Gc(\bfX\cup\bfzero)}}  \phi(R_{\bfzero, 
\bfX})$}}\includegraphics[width = 
2.4in]{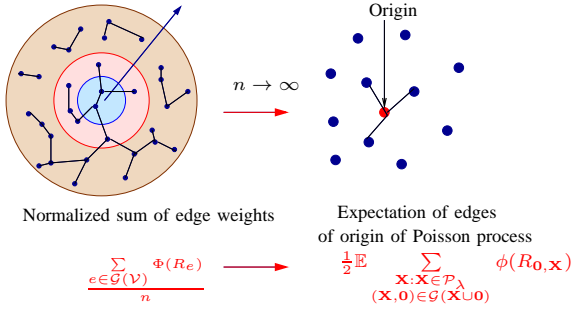}\ep\ec\caption{Pictorial 
representation of LLN for graph functionals of uniform or Poisson 
point sets.}\label{fig:lln}\end{figure}

The law of large numbers for functionals on graphs enables us to 
evaluate the limit\footnote{Nature of convergence is convergence of 
means and complete convergence (c.m.c.c) and implies almost-sure 
convergence.} in (\ref{eqn:D}). This law applies to graphs which are 
random in the sense that the vertex set is a marked random point 
set. LLN on graphs is based on the so-called {\em objective method}. 
Steele \cite{Steele:book}  coined this term for a philosophy 
whereby, loosely speaking, one describes the limiting behavior of 
functionals on finite point sets of binomial process in terms of 
related functionals defined on infinite Poisson point sets. Also see 
Fig.\ref{fig:lln}. Penrose and Yukich 
\cite{Penrose&Yukich:01AAP,Penrose&Yukich:02AAP,Penrose&Yukich:03AAP} 
introduce a concept of {\em stabilizing} functionals and  use the 
objective method to establish a strong law of large numbers for 
graph functionals \cite[P. 287]{Penrose&Yukich:02AAP}. In order to 
apply this law, some conditions need to be satisfied in terms of 
bounded moments. In the lemma below, we place these conditions on 
the correlation function.

\bl[Conditions for LLN]\label{lemma:conditions} The graph functional 
in (\ref{eqn:D}) satisfies the conditions for law of large numbers 
for graph functionals derived in \cite[P. 
287]{Penrose&Yukich:02AAP}, when the correlation function $g$ is 
monotonically non-increasing with the edge-lengths, $g(\infty) =0$, 
and  $g(0) = M <1$. Hence, the graph functional 
in (\ref{eqn:D}) converges almost surely to a constant.
\el

\noindent Proof : See appendix~\ref{proof:conditions}. 

\begin{theorem}[LLN]\label{thm:llnexponent}
Under the conditions stated in Lemma~\ref{lemma:conditions}, for 
nodes placed according to $\Uc_{n,\lambda}$ or $\Pc_{n,\lambda}$, 
with node density $\lambda$ and region area $\frac{n}{\lambda}$, 
from the law of large numbers for graph functionals, the  expression 
for the error exponent $D$ in (\ref{eqn:D}) for Neyman-Pearson 
detection of the GMRF defined by the NNG is given by \beq D= 
\frac{1}{2} \Bigl[\frac{1}{2}\Ebb \!\!\!\!\!\!\sum_{\substack{\bfX: 
\bfX \in \Pc_{\lambda},\\(\bfzero, \bfX)\in \snng(\bfX\cup\bfzero)}}  
\!\! \!\!\!\!  f(g(R_{\bfzero, \bfX}))+\log K +\frac{1}{K}-1  
\Bigr],\label{eqn:Dllnexponent}\eeq where\beq f(x)\defeq 
\log[1-x^2]+ \frac{2 x^2}{K[1-x^2]}, \label{eqn:f}\eeq $K$ is the 
ratio of variances defined in (\ref{eqn:K}), and $R_{\bfzero,\bfX}$ 
are the (random) lengths of edge $(\bfX,\bfzero)$ incident on the 
origin in a NNG, when the nodes are distributed according to 
homogeneous Poisson process $\Pc_\lambda$, of intensity $\lambda$.
\end{theorem}

\noindent Proof :  Apply LLN to (\ref{eqn:D}). See 
appendix~\ref{proof:llnexponent}.  \qed 

In the theorem above, the law of large numbers yields the same 
limit\footnote{In general, the limit is not the same for Poisson and 
binomial processes.  For a different problem, we show that the error 
exponents are affected by a random sample size 
\cite{Anandkumar&Tong:06ICASSP}.} under the Poisson or the binomial 
process. Thus, we provide a single expression for the error exponent 
under both the processes. Also, the above theorem provides the error 
exponent in terms of the expectation of a graph functional around 
the origin, with the points drawn from an infinite Poisson process.  
Thus, the functional is reduced to a localized effect around the 
origin. This is an instance of the broad concept of {\em 
stabilization} which states that the local behavior of the graph in 
a bounded region is unaffected by points beyond a finite (but 
random) distance from that region. NNG is one such stabilizing graph 
with translation and scale-invariance \cite[Lemma 
6.1]{Penrose&Yukich:01AAP}.



\begin{figure*}[t]
\subfloat[a][Different values of correlation coefficient $a$, nugget 
$M=0.5$.]{
\begin{minipage}{3.5in}
 \centering
\begin{psfrags}
\psfrag{K in db}[l]{\scriptsize $K$ in dB}\psfrag{Error exponent 
D}[l]{\scriptsize Error Exponent $D$} \psfrag{A val 
a1}[l]{\scriptsize $a=0$}\psfrag{A val a3}[l]{\scriptsize $a \to 
\infty$}\psfrag{A val a2}[l]{\scriptsize 
$a=0.5$}\includegraphics[height=1.8in]{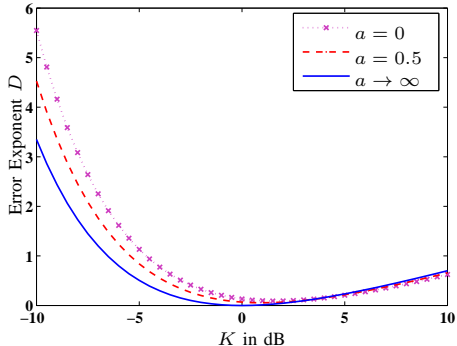}\end{psfrags}
\end{minipage}}\hfil 
\subfloat[b][Different values of nugget $M$, correlation coefficient 
$a=0.5$.]{
\begin{minipage}{3.5in}\centering
\begin{psfrags}
\psfrag{K in db}[l]{\scriptsize $K$ in dB}\psfrag{Error exponent 
D}[l]{\scriptsize Error Exponent $D$} \psfrag{M val 
m1}[l]{\scriptsize $M=0$}\psfrag{M val m3}[l]{\scriptsize 
$M=0.9$}\psfrag{M val m2}[l]{\scriptsize 
$M=0.5$}\includegraphics[height=1.8in]{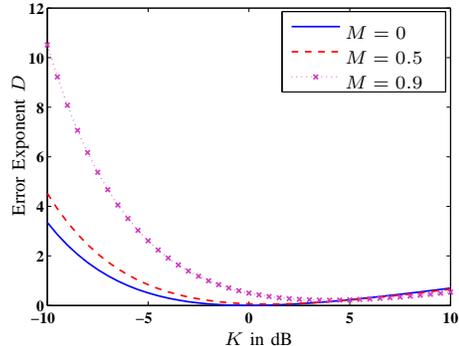}\end{psfrags}
\end{minipage}}
\caption{Error exponent $D$ vs. ratio of variances $K$, node density 
$\lambda =1$. See  
(\ref{eqn:dzero}-\ref{eqn:expcorr}).}\label{fig:dzeroinf} 
\end{figure*}

\subsection{Explicit Form for Nearest-Neighbor Graph}\label{subsec:explicit_results}

The evaluation of the expectation of the edge functional  in 
(\ref{eqn:Dllnexponent}) is complicated and needs further 
simplification. In order to obtain the final form of the exponent, 
we exploit some special properties of the NNG. It turns out that the 
expectation term is easier to evaluate for the directed 
nearest-neighbor graph rather than the undirected version. We 
therefore split the sum of edge functionals in (\ref{eqn:D}), using 
the fact that the directed NNG counts the weights from biroots or 
mutual neighbors twice, while the undirected version counts only 
once.  See Fig.\ref{fig:directed_graph}. We therefore split the sum 
of the edge functionals of the undirected NNG as \beqn\nn 
 \sum_{e \in \snng(\V)}  f(g(R_e))&=& \sum_{e 
\in \sdnng(\V)}  f(g(R_e)) \\&&{-}\: \frac{1}{2}  \sum_{e \in 
\sbirootnng(\V)}  f(g(R_e)),\label{eqn:directed} \eeqn where 
$\nng(\V)$, $\dnng$ and $\birootnng\subset \dnng$ are the undirected 
NNG, the directed NNG, and edges between the biroots or the mutual 
neighbors of the directed NNG, respectively.  Now, we evaluate the 
expectation for the two terms separately, since expectation is 
linear. A similar approach is employed in \cite{Wade:AAP07}. 

We now provide an expression for the limit of the edge functional 
based on the distribution of distances of the directed NNG, which 
are related to hitting or vacancy probabilities of the spatial point 
process, which are typically exponential or gamma distributed, 
similar to their one-dimensional counterparts \cite{Baddeley:book}.

\bl[Expectation of Edge Functional]\label{lemma:edge} The 
expectation term of the edge functional in (\ref{eqn:Dllnexponent}) 
is given by\beq \frac{1}{2}\Ebb 
\!\!\!\!\!\!\!\!\!\!\sum_{\substack{\bfX: \bfX \in 
\Pc_{\lambda},\\(\bfzero, \bfX)\in \snng(\bfX\cup\bfzero)}}  
\!\!\!\!\!\!\!\!\!\! f(g(R_{\bfzero, \bfX}))=  \Ebb f(g( Z_1))- 
\frac{\pi}{2 \omega}\Ebb f( g(Z_2)),\label{eqn:edge}\eeq where  
$Z_1$ and $Z_2$ are Rayleigh distributed with variances $(2 \pi 
\lambda)^{-1}$ and $(2 \omega \lambda)^{-1}$,  and $\omega$ is given 
by  \beq \omega = \frac{4 \pi}{3} + \frac{\sqrt{3}}{2} \approx 
5.06,\label{eqn:omega}\eeq and is the area of the union of two unit- 
radii circles with centers unit distant apart. \el

\noindent Proof :  See appendix~\ref{proof:edge}.  \qed 

In the theorem below, we combine Lemmas~\ref{lemma:conditions}, 
\ref{lemma:edge}, and Theorem~\ref{thm:llnexponent} to obtain  the 
final form of the error exponent.

\begin{theorem}[Expression for $D$]\label{thm:exponent}
Under the assumptions (\ref{eqn:selfcorr}-\ref{eqn:corr}) and 
conditions stated in Lemma~\ref{lemma:conditions}, for a GMRF with 
NNG dependency and correlation function $g$ and nodes drawn from the 
binomial or the Poisson process with node density $\lambda$ and 
region area $\frac{n}{\lambda}$, the error exponent $D$ for 
Neyman-Pearson detection is \beqn D_g(K,M,\lambda)&{}={}& 
\frac{1}{2} \bigl[ \Ebb f(g( Z_1),K)- \frac{\pi}{2 \omega}\Ebb f( 
g(Z_2),K)\nn\\&&{+}\:\log K +\frac{1}{K}-1 
\bigr],\label{eqn:Dexpectation}\eeqn where\beq f(x,K)\defeq 
\log[1-x^2]+ \frac{2 x^2}{K[1-x^2]}.\eeq  $Z_1$ and $Z_2$ are 
Rayleigh distributed with second moments $(2 \pi \lambda)^{-1}$ and 
$(2 \omega \lambda)^{-1}$.  \end{theorem}


The above theorem holds for any general correlation function. In 
(\ref{eqn:Dexpectation}), except for the first two $f$-terms which  
capture the  correlation structure of the GMRF, the remaining terms 
represent the detection error exponent for two IID Gaussian 
processes. In the corollary below, we specialize 
(\ref{eqn:Dexpectation}) to the case of constant correlation.  In 
this case, the two $f$-terms reduce to a single term.

\begin{corollary}[Constant Correlation] For constant values of the  
correlation, the error exponent $D$ is independent of the node 
density $\lambda$ and   \ben 
\item for constant positive correlation or $g(R_e) \equiv M<1,\,\forall e \in \E,$ we have \beqn 
D(K,M)&{}={}&\frac{1}{2} \Bigl[\log K+\frac{1}{K}-1 
\nn\\&&{+}\:(1-\frac{\pi}{2\omega}) f(M,K) 
\Bigr],\label{eqn:dzero}\eeqn where $f$ and $\omega$ are given by 
(\ref{eqn:f}) and (\ref{eqn:omega}). 
\item for the independent case or $g(R_e) \equiv 0,\,\forall e \in \E,$ we have \beq D(K,0) =\frac{1}{2} \Bigl[\log K +\frac{1}{K}-1 
\Bigr].\label{eqn:dinf}\eeq \een\end{corollary}

In the above corollary, we verify that (\ref{eqn:dzero}) reduces to  
(\ref{eqn:dinf}), on substituting $M=0$. In (\ref{eqn:dzero}), the 
effect of correlation can be easily analyzed through the sign of the 
function $f(M,K)$.  Also, \bsbcase{f(M,K)} <0,& for 
$K>\frac{2}{1-M^2}$,\\ >0, & for  $K<2$.\esbcase Therefore,   at 
large variance-ratios, the presence of correlation hurts the 
asymptotic performance, when compared with the independent case. But 
the situation is reversed at low values of the variance ratio and 
the presence of correlation helps in detection performance. In the 
next section, we will draw similar conclusions when the correlation 
function is the exponential function through numerical evaluations.
\subsection{Numerical Results}\label{subsec:numerical}
In this section, we focus on a specific correlation function namely 
the exponential-correlation function, \beq g(R) = 
M\myexp^{-aR},\,\,a>0 , 0<M<1.\label{eqn:expcorr}\eeq  Using 
Theorem~\ref{thm:exponent}, we numerically evaluate $D$ through 
Monte-Carlo runs.  In (\ref{eqn:Dexpectation}), the error exponent 
is an implicit function of the correlation coefficient $a$, through 
the correlation function $g$. For fixed values of $K$ and $M$, we 
have \beq D(K,M, \lambda,a) = D(K,M, 
1,\frac{a}{\sqrt{\lambda}}),\eeq which we obtain by changing the 
integration variable in the expectation term in 
(\ref{eqn:Dexpectation}).  Therefore,   in terms of the error 
exponent, increasing the node density $\lambda$ is equivalent to a 
lower correlation coefficient at unit density.  Here, we plot only 
the effects of correlation coefficient $a$ and nugget $M$ on $D$.

In Fig.\ref{fig:dzeroinf}(a), we plot the error exponent at 
$\lambda=1$ and $M=0.5$, for different values of correlation 
coefficient $a$.  Note, the cases $a=0$ and $a\to \infty$ correspond 
to (\ref{eqn:dzero}) and (\ref{eqn:dinf}). We notice that a more 
correlated GMRF or the one with smaller $a$, has a higher exponent 
at low value of $K$, whereas the situation is reversed at high $K$. 
Equivalently, increasing the node density $\lambda$ improves the 
exponent at low value of $K$, but not at high $K$. Also, when the 
variance ratio $K$ is large enough,  $D$ appears to increase 
linearly with $K$ (in dB), and the correlation coefficient $a$ and 
nugget $M$ appear to have little effect, as expected from Theorem 
\ref{thm:exponent}. In Fig.\ref{fig:dzeroinf}(b), we plot the 
exponent at constant correlation coefficient $a=0.5$ for different 
values of the nugget $M$. Also note, $M=0$ reduces to the 
independent case. We notice a similar behavior as the correlation 
coefficient.  A higher value of $M$ results in a higher exponent at 
low $K$, but not at high $K$.  
\section{Conclusion}\label{sec:conclusion}

In general, finding the closed form  detection error exponent is not 
tractable. The graphical structure of the Markov random field allows 
us to exploit  existing results in spatial probability literature. 
We  employed the law of large numbers for graph functionals to 
derive the detection error exponent for a Gauss-Markov random field 
with  nearest-neighbor dependency graph. We then investigated  the 
influence of model parameters such as the variance ratio and the 
correlation function on the  error exponent.  

In this paper, we have assumed identical variance at every sensor. 
However, a spatially varying SNR model can be incorporated into our 
results.    We have focused on the GMRF defined by the acyclic 
dependency graph and derived the exponent for the nearest-neighbor 
graph. This is a simplifying assumption. Although, the law of large 
numbers is valid for a number of proximity graphs, which have edges 
between ``nearby" points, the actual evaluation of the 
log-likelihood ratio and the exponent are intractable for most of 
these graphs. We have not considered correlation under null 
hypothesis for which one requires a LLN with correlated marks. We 
have also not considered the case when the signal field is not 
directly observable, resulting in a hidden GMRF.  The sparse 
structure of the potential matrix is no longer valid under such a 
scenario. However, note, GMRF with small neighborhood has been 
demonstrated to approximate the hidden GMRF 
\cite{Rue&Steinsland:04Stat} as well as Gaussian field with long 
correlation lengths \cite{Rue&etal:02Stat}, reasonably well. 

The error exponent can be employed as a performance measure for 
network design. In \cite{Anandkumar&Tong&Swami:07SPsub}, we utilize 
the closed form derived in this paper to obtain an optimal node 
density that maximizes the exponent subject to a routing energy 
constraint.  We have also proposed minimum energy  data fusion and 
routing schemes that exploit the correlation  structure of Markov 
random field in a related publication 
\cite{Anandkumar&etal:08INFOCOM}. We further investigate tradeoffs 
between the routing energy and the resulting error exponent in 
\cite{Anandkumar&etal:08ISIT}.


\section*{Acknowledgment}
\noindent  The  authors thank Prof. J.E. Yukich for extensive 
discussions and clarifications regarding the law of large numbers 
for graph functionals. The authors also thank the anonymous 
reviewers and the associate editor Prof. A. Host-Madsen for detailed 
comments that substantially improved this paper.

\begin{appendix}

\subsection{Proof of 
Theorem~\ref{thm:potential&determinant}}\label{proof:potential&determinant}

\noindent Using the expression $\bfA\Sigmabf=\bfI$, we have the 
following identities: \beqn A(i,i) &+& \sum_{j \in \nbd(i)} A(i,j) 
\frac{\Sigma(i,j)}{\Sigma(i,i)}= 
\frac{1}{\Sigma(i,i)},\label{eqn:aiiexp_gen}\\ A(i,i) &+& 
A(i,j)\frac{\Sigma(j,j)}{\Sigma(i,j)} + \sum_{\substack{k \in 
\nbd(i)\\k \neq j}} A(i,k)\frac{\Sigma(j,k)}{\Sigma(i,j)} 
\nn\\&&\qquad\qquad\qquad\qquad\quad=0,\,\forall j \in \nbd(i), 
\label{eqn:aijexp_gen}\eeqn  where (\ref{eqn:aiiexp_gen}) is 
obtained by the sum-product of $i^{\tha}$ row and $i^{\tha}$ column 
of $\bfA$ and $\Sigmabf$. Similarly, (\ref{eqn:aijexp_gen}) is 
obtained by sum-product of $i^{\tha}$ row of $\bfA$  and $j^{\tha}$ 
column of $\Sigmabf$ and  dividing by $\Sigma(i,j)$.  In 
(\ref{eqn:aijexp_gen}), by  acyclicity for $k \in \nbd(i)$ and $k 
\neq j$, we have $j \nsim k$. From (\ref{eqn:nonneighbor_gen}), we 
have \[\frac{\Sigma(j,k)}{\Sigma(i,j)} 
=\frac{\Sigma(i,k)}{\Sigma(i,i)},\quad \forall\,j,k \in \nbd(i),k 
\neq j. \]Subtracting (38) from (37), only the terms with $A(i,j)$ 
survive and hence, we obtain $A(i,j)$. Substituting all the 
$A(i,j)$'s in  (\ref{eqn:aiiexp_gen}), we obtain $A(i,i)$. Hence, 
all the coefficients of potential matrix $\bfA$ are given by 
(\ref{eqn:aij_general}). 

Let $|\bfA^{(n)}|$ be the determinant of the potential matrix of $n$ 
nodes. Assume $n>1$, since we have $|\bfA^{(1)}|=\Sigma(1,1)^{-1}$.  
The determinant of the potential matrix is the product of 
determinants of the connected components.  We therefore consider 
only  one component $\Gc'(\V',\E') \subseteq \Gc$. Assume $\Gc'$ has 
at least one edge, otherwise we have for diagonal matrix 
$|\bfA^{(n)}| = \prod_{i \in \V'} \Sigma(i,i)^{-1}$. Since $\Gc'$ is 
acyclic, it has a leaf, \ie  there is some vertex $a$ with degree 
$1$. Let $b$ be its only neighbor. We assume the vertices have been 
ordered $\V'=\{V_1,\ldots,V_n\}$ so that $V_{n-1} = b, V_n = a$.  
Then $\bfA^{(n)}$ has the following form \[ \bfA^{(n)} = 
\begin{bmatrix} \cdot & \cdots & \cdot & \cdot&0\\ \vdots&\vdots& 
\vdots&\vdots&\vdots\\ \cdot&\cdots&\cdot&\cdot&0\\\cdot&\cdots&\cdot&A(n-1,n-1)&A(n-1,n)\\
0&\cdots&0&A(n,n-1)&A(n,n)\end{bmatrix},\] where we have from 
(\ref{eqn:aij_general}), {\scriptsize \beqnn A(n,n) \!\! \!\! &=& 
\!\!\!\!  \frac{\Sigma(n-1,n-1)}{[\Sigma(n,n) \Sigma(n-1,n-1) - 
\Sigma(n,n-1)^2]},\\ A(n-1,n)\!\! \!\!  &=& \!\! \!\! 
\frac{-\Sigma(n,n-1)}{[\Sigma(n,n) \Sigma(n-1,n-1) - 
\Sigma(n,n-1)^2]},\\ A(n-1,n-1) \!\!\!\!  &=&\!\!\!\!   
\frac{1}{\Sigma(n-1,n-1)} - A(n-1,n) 
\frac{\Sigma(n,n-1)}{\Sigma(n-1,n-1)} + C,\eeqnn }where $C$ 
represents contributions from nodes in $\V'\backslash V_n$  \ie  
with node $V_n$ removed, and having an edge with $V_{n-1}$. 
Multiplying the $n^{\tha}$ column by 
\[\frac{A(n,n-1)}{A(n,n)}=\frac{-\Sigma(n,n-1)}{\Sigma(n-1,n-1)}\] and subtracting it from $(n-1)^{\tha}$ 
column and   using the determinant rule, we have  \beq|\bfA^{(n)}| = 
\begin{vmatrix} \cdot & \cdots & \cdot & \cdot&0\\ \vdots&\vdots& 
\vdots&\vdots&\vdots\\ \cdot&\cdots&\cdot&\cdot&0\\\cdot&\cdots&\cdot&A'(n-1,n-1)&A(n-1,n)\\
0&\cdots&0&0&A(n,n)\end{vmatrix},\label{eqn:alteredmatrix}\eeq where 
\beqn\nn A'(n-1,n-1)\!\!\!\! &\defeq&\!\!\!\!\! 
A(n-1,n-1)\\&&{\!\!\!\!+}\:  \frac{\Sigma(n,n-1)}{\Sigma(n-1,n-1)} 
A(n,n-1).\label{eqn:A'}\eeqn  Hence, we have\[|\bfA^{(n)}|= 
A(n,n)|\bfM_n|,\quad\mbox{for} \quad n>1,\] where $\bfM_n$ is the 
minor of $A(n,n)$ in  (\ref{eqn:alteredmatrix}).  Substituting in 
(\ref{eqn:A'}), we have $A'(n-1,n-1)=C$, where as noted before, $C$ 
is the contributions from nodes in $\V'\backslash V_n$  and having 
an edge with $V_{n-1}$. This implies that $A'(n-1,n-1)$ is the 
coefficient in the potential matrix  for the subgraph induced by 
$\V' \backslash V_n$.  Since only    $ V_{n-1}$ has an edge with 
$V_n$, coefficients of  nodes other than  $V_n$ and $V_{n-1}$ are 
unaffected by the removal of $V_n$. Hence,  $\bfM_n$ is the 
potential matrix    for the subgraph induced by $\V' \backslash 
V_n$,  
\[ \bfM_n=\bfA^{(n-1)}.\]Since $\V' \backslash V_n$ is acyclic, a 
leaf is always present, rearrange the rows such that $\bfA^{(n-1)}$ 
has a leaf in the last two rows, \ie  it has the same structure as 
in (\ref{eqn:alteredmatrix}).  Remove a leaf in each step of the 
recursion, until all the edges are removed, then find the 
determinant with the diagonal matrix consisting of the remaining 
nodes and we obtain (\ref{eqn:determinant_general}). \qed




\subsection{Proof of 
Lemma~\ref{lemma:conditions}}\label{proof:conditions}

We can regard $Y_i$'s as marking, since under $\Hc_0$ they are 
i.i.d. independent of spatial point process. The strong 
stabilization condition is satisfied for NNG \cite[P. 1023, Lemma 
6.1]{Penrose&Yukich:01AAP}. We therefore only need to prove the 
uniform bounded moment condition. We express the edge functional as 
the sum of two functionals, for $i \sim j$, given by \beqn 
\phi_1(\Rij) &\defeq&  
-\log[1-g^2(\Rij)],\label{eqn:phi1}\\\phi_2(\Rij) &\defeq& 
\frac{g^2(\Rij)[Y_i^2+ Y_j^2]- 2 g(\Rij) Y_i Y_j}{1-g^2(\Rij)}. 
\label{eqn:phi2}\eeqn Given a finite marked set $\Vc$, the sum 
functional is denoted by $H$ \ie  \beq H_k(\Vc) 
\defeq \sum_{\substack{(i,j) \in \snng(\V)\\i,j \in \Vc}} \frac{\phi_k(\Rij)}{2},\,\, k=1,2.\eeq 
Given $H_k$, we denote the {\em add one cost} 
\cite[(3.1)]{Penrose&Yukich:02AAP}, which is the increment in $H$, 
caused by inserting a marked point at the origin into a finite 
marked set $\Vc$, by \beq \Delta_k(\Vc) \defeq H_k(\Vc \cup 
\{\bfzero\}) - H_k(\Vc).\eeq


$H_1$ satisfies the polynomial-bounded condition 
\cite[(3.3)]{Penrose&Yukich:02AAP}, since  $\phi_1$ in 
(\ref{eqn:phi1}) is a  
 finite function,  and the number of edges in  NNG is at most $n-1$, 
for $n$ points. However,  the functional $H_2$  does not  satisfy 
the polynomial-bounded condition   since the measurements $Y_i$ in 
(\ref{eqn:phi2})  are unbounded. Instead, we define truncated random 
variable $Z$ as \bsbcase{ Z\defeq} Y , & if $|Y| \leq C \log n$,\\ 
\mbox{sgn}(Y) C \log n, & o.w,\esbcase where $\mbox{sgn}$ is the 
sign function and $C>0$ is a constant. Consider the functionals 
$H'_2, \phi_2'$ by replacing $\bfY_n$ with $\bfZ_n$ in $H_2$ and 
$\phi_2$ respectively. Now, $H'_2$ is polynomially bounded. Further, 
we have $\lim_{n\to \infty} Z \overset{\as}{\to} Y$ and hence, 
$\lim_{n\to \infty} (H'_2 - H_2) \overset{\as}{\to} 0$.

\bd[Uniform Bounded Moments for $\phi_i$]Define $\Uc_{m,A}$ to be 
$m$ uniform random variables on $A \in \Bc$ and $R_{\bfzero,\bfX}$ 
to be the (random) lengths of the edge $(\bfzero,\bfX)$ in graph 
$\Gc$ incident on the origin. Then, the bounded $p$-moment condition 
\cite[(3.7)]{Penrose&Yukich:02AAP} \beqn\sup \limits_{A \in \Bc , 0 
\in A} \sup \limits_{m \in 
[\frac{\lambda|A|}{2},\frac{3\lambda|A|}{2} ]} \Ebb  
[\sum_{\substack{(\bfzero,\bfX)\in \Gc\\\bfX \in \Uc_{m,A}}} 
\phi_k(R_{\bfzero,\bfX})]^p\nn\\ < \infty, 
k=1,2,\label{eqn:uniform_moment}\eeqn is true for some $p\geq 1$.\ed

Without the above condition, nothing can be said about the almost 
sure convergence, although, by Fatou's lemma, the limit of the LLN 
would be a bound on $D$.  



Since $\phi_1$ and $\phi_2$ are  decreasing functions edge length, 
with maximum at zero, we have\[ 
\Ebb[\!\!\!\!\sum_{\substack{(\bfzero,\bfX)\in \snng(\bfX)\\\bfX \in 
\Uc_{m,A}}}\!\!\!\! \phi_k(R_{\bfzero,\bfX})]^p< C^p 
\Ebb[\phi_k(0)]^p,\,\, k=1,2, \,\,\forall p >0,\]  where $C$ is the 
kissing number, a constant, and $\Deg(0) \leq C$ for the NNG. Now, 
$\phi_1(0)= -\log[1-M^2]<\infty$, since $g(0)=M<1$, and 
\[\Ebb[\phi_2(0)^p]<  \frac{M^p}{(1-M^2)^p}\Ebb[Y_i-Y_j]^{2p}<\infty,\] since 
$Y_i,Y_j \overset{i.i.d.}{\sim} \Nc(0,\sigma_0^2)$. Hence, the 
uniform-bounded moment for $\phi_k$ in (\ref{eqn:uniform_moment}) 
holds. 

Now,   we  show the uniform-bounded moment for $H$ 
\cite[(3.2)]{Penrose&Yukich:02AAP}, obtained by replacing $\phi_k$ 
in (\ref{eqn:uniform_moment}) by $\Delta_k$. The positive part of 
$\Delta_k$ is bounded by $\Deg(0) \phi_k(0)$, whose expectation is 
shown to be finite. For the negative part  
$\Delta_k(\Uc_{m,\Ac})^-$,  along the lines of \cite[Lemma 
6.2]{Penrose&Yukich:01AAP}, let $\bfone\{\nnbd(i)=0\}$ be the event 
that the origin is the nearest neighbor of $i \in \Uc_{m,\Ac}$. 
Then, the number of deleted edges on adding the origin is   given by 
$\sum_{i=1}^m \bfone\{\nnbd(i)=0\}\leq C$, we have  
$\Delta_k(\Uc_{m,\Ac})^-\leq C \phi_k(0)$, whose expectation is 
shown to be finite. Hence, the bounded-moment condition for $H$ 
holds and LLN is applicable. \qed

\subsection{Proof of Theorem~\ref{thm:llnexponent}}\label{proof:llnexponent}
\noindent We have the distribution of $\bfY_n$ under the null 
hypothesis \beq \nn \Pb[\bfY_n|\V;\Hc_0] = \frac{1}{(2 \pi 
\sigma_0^2)^{\frac{n}{2}}}\exp{\Bigl( - \frac{\sum_{i=1}^n Y_i^2}{2 
\sigma_0^2} \Bigr)}.\eeq \noindent    Therefore,   the limit of the 
determinant is given by \beq \lim_{n \to \infty} 
\frac{\log|\Sigma_{0,\V}|}{2n} = \log \sigma_0.\eeq       We have 
$\sum_{i=1}^n \frac{Y_i^2}{n}\to \Ebb[Y_1^2;\Hc_0] = \sigma_0^2$ \as 
under $\Hc_0$. Therefore,   the term in (\ref{eqn:D})  
\beq\nn\sum_{i=1}^n\Bigl(\frac{1}{\sigma_1^2}-\frac{1}{\sigma_0^2}\Bigr)\frac{Y_i^2}{n} 
\to \Bigl(\frac{\sigma_0}{\sigma_1}\Bigr)^2-1.\eeq       

\noindent By Lemma~\ref{lemma:conditions}, the conditions for LLN 
hold and therefore as $n \to \infty$, \beqn -\frac{1}{n} 
\sum\limits_{e \in \snng(\V)}  \log[1-g^2(R_e)] \qquad \qquad \qquad 
\qquad \qquad&{}&\nn\\ \to -\Ebb \sum_{\substack{\bfX:\bfX \in 
\Pc_\lambda\\(\bfzero, \bfX)\in \snng(\bfX)}}\log[1-g^2(R_{\bfzero, 
\bfX})],\\ \frac{1}{n} \sum\limits_{\substack{(i,j) \in 
\snng(\V)\\i<j}} \frac{g^2(\Rij)[Y_i^2+ Y_j^2]- 2 g(\Rij) Y_i 
Y_j}{[1-g^2(\Rij)]\sigma_1^2}\nn\\ \to \Ebb 
\sum_{\substack{\bfX:\bfX \in \Pc_\lambda\\(\bfzero, \bfX)\in 
\snng(\bfX)}}\frac{2g^2(R_{\bfzero,\bfX})}{1-g^2(R_{\bfzero, 
\bfX})}\Bigl(\frac{\sigma_0}{\sigma_1}\Bigr)^2,\label{eqn:sumy}\eeqn   
where, in (\ref{eqn:sumy}) we first take the expectation over 
$Y_i$'s  and use the fact that $\Ebb[\frac{Y_\bfzero^2+ 
Y_\bfX^2}{\sigma_1^2}] = 2(\frac{\sigma_0}{\sigma_1})^2$ and 
$\Ebb[Y_\bfzero Y_\bfX]=0$. Collecting all the terms we have 
(\ref{eqn:Dllnexponent}). \qed


%

\subsection{Proof of Lemma~\ref{lemma:edge}}\label{proof:edge}

\begin{figure}[t]\bc\bp\psfrag{0}[r]{\scriptsize $\bfzero$}\psfrag{nn(0)}[l]{\scriptsize $\nnbd(\bfzero)$}
\psfrag{1}[c]{\scriptsize $Z_1$}
\includegraphics[width=1.5in]{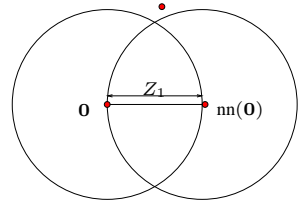}\ep\ec
\caption{Illustration of the event that the origin is a biroot in 
the directed NNG. This implies that there is no other point in the 
union of the circles shown above. See 
(\ref{eqn:biroot}).}\label{fig:biroot}\end{figure}

We use an approach similar to \cite{Wade:AAP07}. Let $B_z(\bfX)$ 
denote a circle of radius $z$, centered at $\bfX$. We take 
expectation on both sides of (\ref{eqn:directed}) for graphs  over 
all the Poisson points $\bfX\cup \bfzero$. Let $\nng(\V)$, $\dnng$ 
and $\birootnng\subset \dnng$ be the undirected nearest-neighbot 
graph, the directed nearest-neighbor graph, and edges between the 
biroots or the mutual neighbors of the directed nearest-neighbor 
graph. See Fig.\ref{fig:directed_graph}.

\vspace{-1em}

 {\small
 \beqn\nn \Ebb[\!\!\!\!\sum_{\substack{\bfX:\bfX\in \Pc_\lambda\\(\bfzero,\bfX)\in 
\snng(\bfX)}} f(g(R_{\bfzero,\bfX}))] \!\!   =  \!\!   
\Ebb[\sum_{\substack{\bfX:\bfX\in \Pc_\lambda,\\(\bfzero,\bfX)\in 
\sdnng(\bfX)}} f(g(R_{\bfzero,\bfX}))]
\\-\frac{1}{2}\Ebb[ \!\!  \!\! \sum_{\substack{\bfX:\bfX\in \Pc_\lambda\\(\bfzero,\bfX)\in 
\sbirootnng(\bfX)}} f(g(R_{\bfzero,\bfX}))].\label{eqn:terms}\eeqn}

\noindent The first term on the right-hand side in (\ref{eqn:terms}) 
simplifies   as \beq  \Ebb[\sum_{\substack{\bfX:\bfX\in 
\Pc_\lambda,\\(\bfzero,\bfX)\in \sdnng(\bfX)}} 
f(g(R_{\bfzero,\bfX}))] = \Ebb[f(g(Z_1))],\eeq where $Z_1$ is the 
unique directed nearest-neighbor distance of the origin with points 
distributed according to  $\Pc_\lambda$, the Poisson point process  
of intensity $\lambda$ on $\Re^2$.  The random variable $Z_1$ is 
like a waiting time, and can be visualized as the time taken for an 
inflating circle to first touch a point from the Poisson process. We 
therefore have $Z_1 
> z$ iff. $B_z(\bfzero)$ does not contain any points from the Poisson 
process, \ie  \beq \Pbb[Z_1 
> z] = \Pbb[\nexists \bfX \neq  \bfzero\in B_z(\bfzero)\cap \Pc_\lambda]= \myexp^{-\lambda \pi z^2}.\label{eqn:Z1}\eeq Therefore,    
$Z_1$ is Rayleigh  with second moment $(2 \pi \lambda)^{-1}$. 

Similarly, for the second term, we need to find the PDF of  the 
nearest-neighbor distance of the origin when the origin is a biroot 
or a mutual nearest neighbor. This event occurs when  the union of 
the circles centered at origin and its nearest neighbor  contains no 
other Poisson point. See Fig.\ref{fig:biroot}. Let $\Ac$ be the 
intersection of the events  that the directed nearest-neighbor 
distance of origin lies in the interval $[z,z+dz]$   and the event 
that origin is a biroot \beqn\nn \Ac&\defeq&(\Pc_\lambda \cap ( 
B_z(\bfzero) \cup B_z(\nnbd(\bfzero))) \backslash 
\{\bfzero,\nnbd(\bfzero)\}=\emptyset)
\\&&\cap (Z_1 \in [z,z+dz])\label{eqn:biroot} .\eeqn Its 
probability  is given by, \beqn \nn\Pbb[\Ac]&=& \Pbb(\mbox{\small 
origin is biroot}|Z_1)\Pbb( Z_1 \in 
[z,z+dz])\\&=&\myexp^{-(\omega-\pi) \lambda z^2} 2\lambda\pi z 
\myexp^{-\lambda\pi z^2}  dz\label{eqn:intersection_prob}\\&=& 
2\lambda\pi z \myexp^{-\omega \lambda z^2} dz=\frac{ 
\lambda}{\omega}[2\omega\pi z \myexp^{-\omega \lambda z^2} 
dz]\\&=&\frac{ \lambda}{\omega} \Pbb(Z_2\in 
[z,z+dz]),\label{eqn:Z2}\eeqn where $\nnbd(\bfzero)$ is the  
nearest-neighbor of the origin and $\omega 
\defeq |B_1(\bfzero)\cup B_1(\mathbf{1})|= \frac{4 \pi}{3} + 
\frac{\sqrt{3}}{2}$, the area of the union of circles unit distant 
apart and $Z_2$  is a Rayleigh variable  with variance $(2 \pi 
\omega)^{-1}$.  Hence, the second term  on the right-hand side in 
(\ref{eqn:terms})  simplifies as \beq \frac{1}{2}\Ebb[   
 \sum_{\substack{\bfX:\bfX\in 
\Pc_\lambda\\(\bfzero,\bfX)\in 
\sbirootnng(\bfX)}}f(g(R_{\bfzero,\bfX}))] = 
\frac{\pi}{2\omega}\Ebb[ f(g(Z_2))].\eeq

From (\ref{eqn:directed}, \ref{eqn:Z1}, \ref{eqn:Z2}), we obtain 
(\ref{eqn:edge}). \qed

\end{appendix}

\begin{biographynophoto}{\bf Animashree 
Anandkumar} (S '02)  received the B.Tech degree in Electrical 
Engineering from Indian Institute of Technology Madras, Chennai, 
India in 2004.  She is currently pursuing her Ph.D. degree in 
Electrical Engineering  at Cornell University, Ithaca, NY. She has 
been a member of the Adaptive Communications and Signal Processing 
Group (ACSP) at Cornell University since August, 2004.

Anima received the Fran Allen IBM Ph.D fellowship for the year 
2008-09,  presented annually to one female Ph.D. student in 
conjunction with the IBM Ph.D. Fellowship Award. She  was named a 
finalist for the Google Anita-Borg Scholarship 2007-08.  She 
received the Student Paper Award  at the 2006 International 
Conference on Acoustic, Speech and Signal Processing  (ICASSP) held 
at Toulouse, France.

Anima Anandkumar's research interests are in the area of 
statistical-signal processing, information theory and networking. 
Specifically, she has been working on detection and estimation, 
asymptotic analysis and in-network function computation, in the 
context of wireless-sensor networks. She has served as a reviewer 
for IEEE transactions on signal processing, IEEE transactions on 
information theory and various IEEE 
conferences.\end{biographynophoto}

\begin{biographynophoto}{\bf Lang Tong} 
(S'87,M'91,SM'01,F'05) is the Irwin and Joan Jacobs Professor in 
Engineering at Cornell University Ithaca, New York.  Prior to 
joining Cornell University, he was on faculty at the West Virginia 
University and the University of Connecticut. He was also the 2001 
Cor Wit Visiting Professor at the Delft University of Technology. 
Lang Tong received the B.E. degree from Tsinghua University, 
Beijing, China, in 1985, and M.S. and Ph.D. degrees in electrical 
engineering in 1987 and 1991, respectively, from the University of 
Notre Dame, Notre Dame, Indiana. He was a Postdoctoral Research 
Affiliate at the Information Systems Laboratory, Stanford University 
in 1991.

Lang Tong is a Fellow of IEEE. He received the 1993 Outstanding 
Young Author Award from the IEEE Circuits and Systems Society, the 
2004 best paper award (with Min Dong) from IEEE Signal Processing 
Society, and the 2004 Leonard G. Abraham Prize Paper Award from the 
IEEE Communications Society (with Parvathinathan Venkitasubramaniam 
and  Srihari Adireddy). He is also a coauthor of five student paper 
awards. He received Young Investigator Award from the Office of 
Naval Research.  

Lang Tong's research is in the general area of statistical signal 
processing, wireless communications and networking, and information 
theory.  He has served as an Associate Editor for the IEEE 
Transactions on Signal Processing, the IEEE Transactions on 
Information Theory, and IEEE Signal Processing Letters.

\end{biographynophoto}

\begin{biographynophoto}{\bf Ananthram Swami} 
received the B.Tech. degree from IIT-Bombay; the M.S. degree from 
Rice University, and the Ph.D. degree from the University of 
Southern California (USC), all in Electrical Engineering.  He has 
held positions with Unocal Corporation, USC, CS-3 and Malgudi 
Systems.  He was a Statistical Consultant to the California Lottery, 
developed a Matlab-based toolbox for non-Gaussian signal processing, 
and has held visiting faculty positions at INP, Toulouse. He is 
currently with the US Army Research Laboratory where his work is in 
the broad area of signal processing, wireless communications, sensor 
and mobile ad hoc networks.

He has served as associate editor of the IEEE Transactions on 
Wireless Communications, IEEE Signal Processing Letters, IEEE 
Transactions on Circuits \& Systems-II, IEEE Signal Processing 
Magazine, and IEEE Transactions on Signal Processing. He was 
co-guest editor of a 2004 Special Issue (SI) of the IEEE Signal 
Processing Magazine (SPM) on `Signal Processing for Networking', a 
2006 SPM SI on `Distributed signal processing in sensor networks', a 
2006 EURASIP JASP SI on Reliable Communications over Rapidly 
Time-Varying Channels', a 2007 EURASIP JWCN SI on `Wireless mobile 
ad hoc networks', and is the Lead Editor for a 2008 IEEE JSTSP SI on 
``Signal Processing and Networking for Dynamic Spectrum Access''.  
He is a co-editor, with Qing Zhao, Yao-Win Hong and Lang Tong, of 
the 2007 Wiley book ``Wireless Sensor Networks: Signal Processing \& 
Communications Perspectives''.

\end{biographynophoto}

{\scriptsize


}

\end{document}